\documentclass[aps,prb,%preprint,
twocolumn, groupedaddress]{revtex4}
\usepackage{mathrsfs}
\usepackage{graphicx}% Include figure files
\usepackage{float}
\usepackage{amsfonts,amssymb,amsmath,mathrsfs}

\begin{document}

\title{Landau-Zener-St\"{u}ckelberg Spectroscopy of a Superconducting Flux Qubit}

\author{Shuang Xu}
\affiliation{National Laboratory of Solid State Microstructures and
Department of Physics, Nanjing University, Nanjing 210093, China}
\author{Guozhu Sun}
\affiliation{Institute of Superconductor Electronics and Department
of Electronic Science and Engineering, Nanjing
University, Nanjing 210093, China\\
Department of Physics and
Astronomy, University of Kansas, Lawrence, KS 66045, USA}
\author{Yang Yu}\email[]{yuyang@nju.edu.cn}
%\homepage[]{Your web page}
%\thanks{}
%\altaffiliation{}
\affiliation{National Laboratory of Solid State Microstructures and
Department of Physics, Nanjing University, Nanjing 210093, China}

\date{\today}

\begin{abstract}
We proposed a new method to measure the energy spectrum of a
superconducting flux qubit. Different from the conventional
frequency spectroscopy, a short triangle pulse is used to drive the
qubit through the anticrossing and generates
Landau-Zener-St\"{u}ckelberg interference patterns, from which the
information of the energy spectrum can be extracted. Without
installing microwave lines one can simplify the experimental setup
and reduce the unwanted effects of noise. Moreover, the method can
be applied to other quantum systems, opening the possibility of
calibrating and manipulating qubits with linear pulses.
\begin{description}\item[PACS numbers]{: 85.25.Cp, 03.67.Lx}\end{description}
\end{abstract}

% insert suggested PACS numbers in braces on next line
% insert suggested keywords - APS authors don't need to do this
%\keywords{}

%\maketitle must follow title, authors, abstract, \pacs, and \keywords
\maketitle

\section {Introduction \label{1}}
It has been shown, both theoretically and experimentally, that
superconducting devices based on Josephson junctions are ideal
architectures successfully displaying macroscopic quantum phenomena
\cite{qrev}. On Josephson junction devices, typical quantum
phenomena like quantum tunneling \cite{cal_leg, tunneling, han2001},
energy level quantization \cite{martinis1985} and coherent
superpositions \cite{nakamura1999, sup, suppcq, cohdyn} have been
predicted and experimentally demonstrated. Temporal evolution of
quantum states and Rabi oscillations have been observed in
macroscopic sense \cite{rabi, cohosc}. By virtue of the convenience
in engineering, many new ideas are expected to be implemented with
Josephson junction devices; one of them is quantum computation
\cite{qrev, qcqi}. The basic requirement of quantum computation is a
quantum bit (qubit). In practice, accurate, efficient and stable
manipulations on qubits are the prerequisites of quantum
computation. People have proved Josephson Junction devices as a
promising candidate for a qubit \cite{qrev, nakamura1999, cohosc,
pcqubit, you}. Furthermore, integrated multi-qubit control has been
attempted with Josephson Junction devices\cite{pashkin,
entangled2qubits, controlcoup, twoqubit, multiqubit}.

In order to control a qubit with high fidelity, first of all, we
have to measure the energy spectrum of the qubit. Conventional
spectroscopy utilizes frequency resonance to measure the absorption
spectrum of the qubit. This method originates from the historical
frequency spectroscopy that has been applied to atoms or molecules
\cite{a1, a2}. In the case of Josephson junction devices, which are
often referred as artificial ``atoms", people inject microwaves to
the junction circuit and measure the absorption spectrum. However,
this method becomes challenging if the energy is over 10 GHz,
because microwaves of this high frequency are hard to optimize with
common electronic devices. Although high frequency microwave sources
are available, they are expensive and the signal becomes noisy after
passing multipliers. Transmission lines and waveguides also limit
the applications due to restricted bandwidth \cite{ieee}. To
overcome this challenge, some groups have developed alternative
techniques such as photon-assisted tunneling \cite{photonat}. In
2008, Berns \emph{et al.} proposed a method called amplitude
spectroscopy \cite{as}. They drove a flux qubit with a microwave of
0.16 GHz, whose amplitude was as large as to sweep through
anticrossings in higher energy levels and produced
Landau-Zener-St\"{u}ckelberg (LZS) interference. They observed
diamond-like interference patterns, from which the information of
energy levels with energies up to 100 GHz was extracted. However,
microwaves have phase indeterminacy, which introduces decoherence of
the qubit and leads to difficulties in exact manipulations. So we
propose another method replacing microwaves with a short triangle
pulse. We have studies the LZS interference generated by a triangle
pulse sweep. With simulations we have obtained the LZS interference
patterns,which encode the information of the spectrum. Remarkably,
this technique is also useful in coherent manipulations on quantum
systems because the signal length can be set within a sufficiently
short time.

This article is organized as follows. In Section \ref{2} we shall
introduce the model in our simulations. Section \ref{3} is devoted
to the results and analytical explanations of a two-level system. In
Section \ref{4} we take a further step to study the case of a
multi-level system. Conclusions are made in Section \ref{5}.

\section {Model \label{2}}

\begin{figure}[h]
\includegraphics[height=5cm]{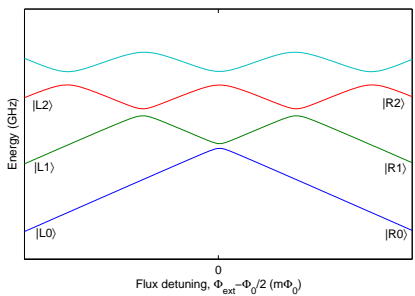}
\caption{\label{spectrum}(color online) Typical energy-level diagram
of a flux qubit, showing the relation between the energy and the
flux detuning $\Phi_{ext}-\Phi_0/2$.}
\end{figure}

The qubit under our discussion is similar to the one in the
experiment by Berns \emph{et al.} \cite{as}. It is a superconducting
flux qubit built with a superconducting loop interrupted by three
junctions. The qubit is biased with an external flux $\Phi_{ext}$
around $\Phi_0/2$, where $\Phi_0\equiv h/2e$ is the flux quantum.
The potential of the qubit has the double-well shape parameterized
by the flux bias. According to quantum mechanics, there are
quantized energy levels located in both wells. If the noises are
suppressed to a sufficiently low degree (much smaller than the
energy gaps), quantum behaviors can be observed. We label the
localized states as $|Ln\rangle$ (left well) and $|Rn\rangle$ (right
well), where $n=0,1,2...$, counting from the ground state. In the
basis composed of the localized states, the Hamiltonian of the qubit
has non-zero off-diagonal elements, which mark the interwell
tunneling. Diagonalizing the Hamiltonian and calculating its
eigenvalues, we obtain the spectrum of the qubit theoretically. The
spectrum is often parameterized by the flux detuning
$\Phi_{ext}-\Phi_0/2$, as drawn in Fig.(\ref{spectrum}). There are
anticrossings at certain values of the flux bias, avoiding energy
degeneracy of the two states in different wells when they are tilted
to the same energy by the external flux bias. In regions other than
the vicinities of the anticrossings, the energy varies linearly with
the flux bias, which is a valid approximation in common cases.

The study of Landau-Zener transitions and St\"{u}ckelberg-type
interference in a qubit has lasted for a long time \cite{as,
oliver2009, LZ4q, la4tls, multilevels, MZI, johansson}. Usually, the
experiments are carried out as follows. The qubit is initialized at
the ground state in a well, e.g. $|L0\rangle$. Then a microwave
signal is injected through the bias line. After some periods of
sweeping, the population in a certain state is measured by using a
dc SQUID juxtaposed aside the qubit to detect the loop current
direction, which is decided by the state of the qubit. The previous
works have observed clear interference patterns and managed to
extract useful data from the patterns. In our simulation, we follow
the same key steps, but have changed some details. First, our sweep
signal is a linear triangle pulse, which can be accurately
programmed and generated. This seems a minor change but brings great
convenience to the experimenters. Moreover, this is closer to the
type of transitions originally studied by Zener so Zener's formula
\cite{Zener} is of higher accuracy in our work. Second, we measure
the system every time when a single pulse ends. This ensures that we
will obtain the direct result of one-turn interference within the
decoherence time of the system. The result has not been disturbed or
diminished by effects of the environment or defects. Therefore, it
is a convenient operation with both accuracy and efficiency.

Now we simulate the revised process of the LZS interference. We
drive the qubit with the time-dependent signal
\begin{equation}\Phi_{ext}(t)=\Phi_0/2+\Phi_i+\text{Trgl}(\Phi_f,\tau,t),\end{equation}
where $\Phi_i$ is the initial flux detuning and
$\text{Trgl}(\Phi_f,\tau,t)$ is the triangle signal parameterized by
$\Phi_f$ and $\tau$, which correspond to the final value of the flux
sweep and the time width respectively(shown in Fig.(\ref{f1}a)). The
explicit expression is
\begin{equation}\text{Trgl}(\Phi_f,\tau,t)=
\left\{\begin{array}{cc}
kt & (0<t<\frac{\tau}{2})\\
k(t-\tau) & (\frac{\tau}{2}<t<\tau) \\
\end{array}\right. ,\end{equation}
where $k$ is the sweep rate
\begin{equation}k=\frac{2(\Phi_f-\Phi_i)}{\tau} \label{k}.\end{equation}
First we consider the two lowest states $|L0\rangle$ and
$|R0\rangle$, that is, to treat the qubit as a pure two-level
system. The reduced Hamiltonian is
\begin{equation} \label{H1}
\hat{H}_{red}=\hbar\left(\begin{array}{cc}
-\Omega(\Phi_{ext}) & \Delta\\
\Delta & +\Omega(\Phi_{ext})\\
\end{array}\right)\ .
\end{equation}
$\pm\Omega(\Phi_{ext})$ are energy frequencies of the ground states
in two wells; $\Delta$ is the tunneling frequency between the two
states. In following text we set $\hbar=1$. When we add signal to
$\Phi_{ext}$, the Hamiltonian is time-dependent. We use density
matrix to calculate the populations in the states. The density
matrix of the qubit is
\begin{equation}
\hat\rho(t)=\left(\begin{array}{cc}
W_{11}(t) & W_{12}(t)\\
W_{21}(t) & W_{22}(t)\\
\end{array}\right)\ .
\end{equation}
$W_{11}$ and $W_{22}$ are populations in $|L0\rangle$ and
$|R0\rangle$ respectively, and obey the unity condition
$W_{11}+W_{22}=1$. $W_{12}$ and $W_{21}$, which are complex
conjugate, mark the coherence of the two states. The time-evolution
of the density matrix satisfies the celebrated Louville equation
\begin{equation}
i\frac{d\hat\rho(t)}{dt}=-i\check{\Gamma}\hat\rho(t)+[\
\hat{H}_{red}, \hat\rho(t)\ ]\ \label{LE}.
\end{equation}
The decaying rate tensor $\check{\Gamma}$ will be ignored because
the signal width is much shorter than the decaying time.

We initialize the qubit on $|L0\rangle$, which means that the
initial condition is
\begin{equation}
\hat\rho(0)=\left(\begin{array}{cc}
1 & 0 \\
0 & 0 \\
\end{array}\right)\ .
\end{equation}
Solve the Louville equation Eq.(\ref{LE}) to obtain the value of
$W_{11}(t)$ at $t=\tau$, the population in $|L0\rangle$ when the
sweep signal ends. Choosing ranges for the signal parameters
$\Phi_f$ and $\tau$, and solving the Louville equation for all
combinations of signal parameters within their ranges, we obtain a
map showing the final $W_{11}$ versus the two parameters.

\section {Results and Analysis \label{3}}

\begin{figure}[hbpt]
\center
\includegraphics[height=5cm]{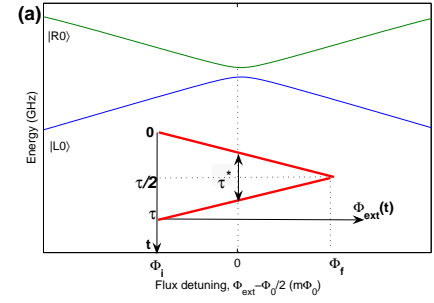}
\includegraphics[height=5cm]{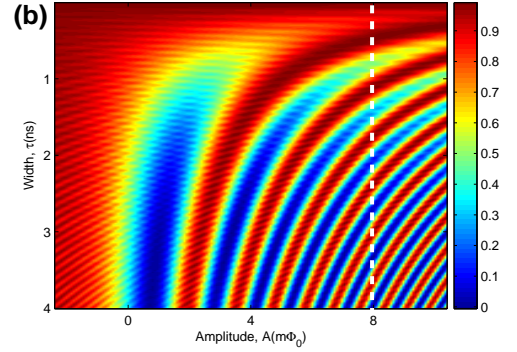}
\includegraphics[height=5cm]{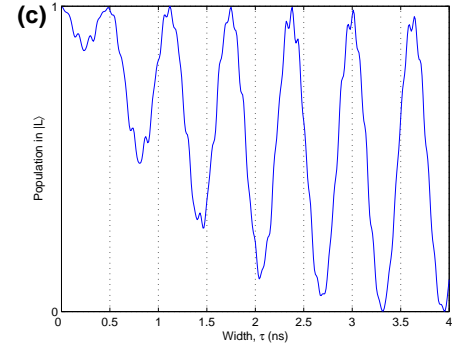}
\includegraphics[height=5cm]{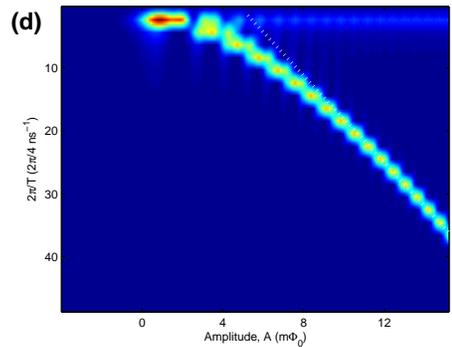}
\caption{\label{f1}(color online) (a)Illustration of the qubit
spectrum. Only the two lowest levels $|L0\rangle$ and $|R0\rangle$
are taken into account. Preset parameters: the slope is 2
GHz/m$\Phi_0$ and $\Delta$ is 2 GHz. The inset shows the details of
the signal, with the time width $\tau$ and the final value $\Phi_f$.
$\Phi_i$ is the initial flux detuning and set to be -5 m$\Phi_0$ in
this simulation. $\tau^\star$ is the interval of two times of
passing through the anticrossing. (b)The simulation result of LZS
interference. The qubit is driven with the signal illustrated in
(a). The population in $|L0\rangle$ begins to oscillate when the
anticrossing is reached. The data at a large amplitude, e.g. along
the white dashed line, can be used to calculate the slope of the
energy spectrum. (c)The data on the white dashed line: population in
$|L0\rangle$ versus the signal width $\tau$ when the amplitude
$\Phi_f$ is fixed at 8 m$\Phi_0$. The period of the oscillation is
about 0.7ns. As a fitting parameter, the energy slope $l\sim$ 2
GHz/m$\Phi_0$, qualitatively agreeing with the common flux qubits.
(d)Discrete one dimensional Fourier transformation of the columns in
(b). Linear relation between $2\pi/T$ and $\Phi_f$ at extremely
large amplitudes is shown. }
\end{figure}

Shown in Fig.(\ref{f1}) are our simulations. In this case, we preset
the spectrum with an energy slope of 2 GHz/m$\Phi_0$ and $\Delta$ of
2 GHz. For the signal we have $\Phi_i=-5 $ m$\Phi_0$, $\tau$ ranging
from 0.01 to 4 ns and $\Phi_f$ from -2 to 10 m$\Phi_0$. The final
population in the initial state $|L0\rangle$ oscillates along both
axes, demonstrating the constructive interference and the
destructive interference. The oscillating periods are determined by
the energy parameters of the qubit, so we can extract the qubit
spectrum from the interference pattern.\\

It is easy to obtain an analytical solution, which can
quantitatively elaborate the result and give specific steps to
extract data. Sweeping through the anticrossing incurs Laudau-Zener
transitions, giving rise to populations in $|R0\rangle$. When
sweeping back to the anticrossing, a phase difference of the two
states $|L0\rangle$ and $|R0\rangle$ is accumulated:
\begin{equation}\varphi=\int_0^{\tau^\star}(\nu_1(t)-\nu_0(t))dt\ \label{int}.\end{equation}
$\nu_0(t)$ and $\nu_1(t)$ are given by the eigenvalues of
Hamiltonian (\ref{H1}):
\begin{equation}\nu_{0,1}(t)=\mp\sqrt{\Omega(t)^2+\Delta^2}\ .
\end{equation}
$\tau^\star$ is the effective width of the signal, equating the
interval of the two times of passing the anticrossing (shown in
Fig.(\ref{f1}a)):
\begin{equation}\tau^\star=\frac{2\Phi_f}{k}=\frac{\Phi_f\tau}{\Phi_f-\Phi_i}
\label{tau} .
\end{equation}
Notice that $\Omega(t)$ varies linearly with time
\begin{equation}
\Omega(t)=l(\Phi_{ext}-\Phi_0/2)=l(kt-\Phi_0/2) (0<t<\tau/2),
\end{equation}
where $l$ is the slope of the spectrum. Substituting it into the
integral Eq.(\ref{int}) and resetting the time parameter, we have
\begin{eqnarray} \varphi &=&
\int_0^{\tau^\star}(\nu_1(t)-\nu_0(t))dt \\
&=& 2 \int_0^{\tau^\star/2}2\sqrt{(lkt)^2+\Delta^2}dt \nonumber\\
&=&
lk\tau^\star\sqrt{\bigl(\frac{\Delta}{lk}\bigr)^2+\bigl(\frac{\tau^\star}{2}\bigr)^2} \nonumber\\
&\ &+2
\frac{\Delta^2}{lk}\ln{\bigl|\frac{lk\tau^\star}{2\Delta}+\sqrt{1+\bigl(\frac{lk\tau^\star}{2\Delta}\bigr)^2}\bigr|}\
\ \label{varphi}.
\end{eqnarray}
Using Eq.(\ref{k}) and Eq.(\ref{tau}) to replace $k$ and
$\tau^\star$ with the original signal parameters, we can write the
above equation as
\begin{eqnarray}
\varphi=&&\frac{\Phi_f\tau}{\Phi_f-\Phi_i}\bigl(\sqrt{\Delta^2+(l\Phi_f)^2} \nonumber\\
&&+\frac{\Delta^2}{l\Phi_f}\ln{\bigl|\frac{l\Phi_f}{\Delta}+\sqrt{1+\bigl(\frac{l\Phi_f}{\Delta}\bigr)^2}\bigr|}\bigr)
\ \label{varphi1}.
\end{eqnarray}
It is this phase difference that causes the interference. It can
proved that the population in the initial state oscillates
sinusoidally with $\varphi$ \cite{Xu}. \\
In the limit of large amplitude driving we have
$$\frac{l\Phi_f}{\Delta}\gg1 ,$$
thus
\begin{equation}
\varphi\approx\frac{l\Phi_f^2}{\Phi_f-\Phi_i}\tau\ ,
\label{phiinlimit}\end{equation} from which $l$ can be extracted as
the fitting parameter.

To test our method, we extract a section in Fig.(\ref{f1}b) along
the white dashed line. This section (see Fig.(\ref{f1}c)) shows the
relation between the population in $|L0\rangle$ and the signal width
$\tau$ at a fixed amplitude that is large enough to validate the
approximation in Eq.(\ref{phiinlimit}). The oscillation period $T$
is about 0.7 ns. According to
Eq.(\ref{phiinlimit}),
$$\frac{l\Phi_f^2}{\Phi_f-\Phi_i}=\frac{2\pi}{T}\sim\frac{2\pi}{0.7}\
 \text{ns}\ ^{-1}.$$
 Note that $\Phi_f$ is fixed at 8 m$\Phi_0$ and
$\Phi_i=-5$ m$\Phi_0$, the slope is 1.83 GHz/m$\Phi_0$. Comparing
with our preset parameter $l=2$ GHz, we reach a precision of at least $90\%$.  \\
In the extremely amplitude regime, Eq.(\ref{phiinlimit}) can be
again approximated as
\begin{equation}\varphi\approx l\Phi_f\tau\
\label{phiinlimit1}.\end{equation}Thus,\begin{equation}\frac{2\pi}{T}\approx
l\Phi_f .\end{equation} To verify this prediction, we extend the
calculation range of $\Phi_f$ and make a discrete one dimensional
Fourier transformation to the interference pattern. The one
dimensional Fourier transformation diagram is shown in
Fig.(\ref{f1}d). We can clearly observe the linear relation at
extremely large values of $\Phi_f$. Actually, when $\Phi_f$ is
larger than 8 m$\Phi_0$, the higher levels should be taken into
account, as is the case discussed in the following section.

\begin{figure}[hbpt]
\center
\includegraphics[height=5cm]{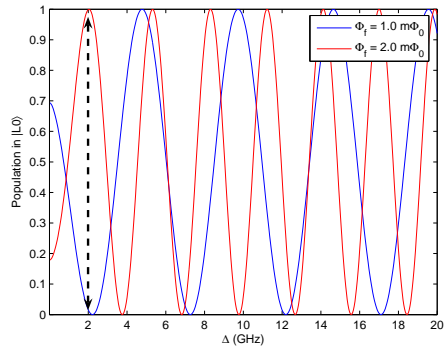}
\caption{\label{deltafit}Plots of the function in Eq.(\ref{pop})
with two sets of parameters. We find that at $\Delta=2\ \text{GHz}$,
both values fit well with the simulation, indicated with the arrow.
Thus we obtain the spectrum parameter $\Delta$.}
\end{figure}

After evaluating $l$, we can make a further step to extract $\Delta$
from Eq.(\ref{varphi1}). We only need to choose two
points in the lower region of Fig.(\ref{f1}b), for instance: \\
$\Phi_f=1.0 \ \text{m}\Phi_0, \tau=3.85\ \text{ns}, \text{Population}=0.00;$\\
$\Phi_f=2.0\ \text{m}\Phi_0, \tau=3.85\ \text{ns}, \text{Population}=0.93.$\\
The population on $|L0\rangle$ is equal to\cite{Xu}
\begin{equation}\frac{1}{2}(1+\cos\varphi),\label{pop}\end{equation} in
which $\varphi$ can be expressed in terms of $\Delta$ as in
Eq.(\ref{varphi1}). So we plot Eq.(\ref{pop}) with $\Delta$ as the
$x$-axis and find the value of $\Delta$ which fits the both points
we have chosen. The result is shown in Fig(\ref{deltafit}). The only
coincident result is $\Delta=2\ \text{GHz}$, which is also precisely
in accordance with our preset parameter.

\section {Multi-level system\label{4}}

\begin{figure}[hbpt]
\center
\includegraphics[height=5cm]{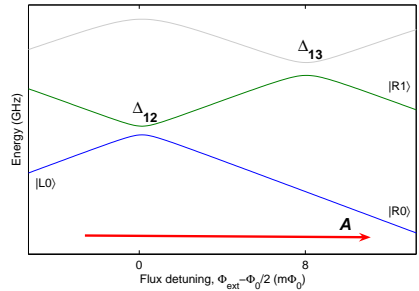}
\caption{\label{3spectrum}(color online) Illustration of the
spectrum of the three levels in question. $\Delta_{12}$ and
$\Delta_{13}$ mark the sizes of the energy gaps at the
anticrossings. The bold arrow illustrates the large amplitudes that
sweep through both anticrossings.}
\end{figure}

If we drive the qubit with a signal whose amplitude is large enough
to reach another anticrossing in higher energy levels, the
interference pattern emerges with interesting characteristics,
especially for some sets of parameters. The simulation approach is
quite similar to the case in Section \ref{3}. We consider the three
states $|L0\rangle$, $|R0\rangle$ and $|R1\rangle$, as shown in
Fig.(\ref{3spectrum}). Under the basis composed of these three
states, the reduced Hamiltonian can be written as
\begin{equation} \label{H}
\hat{H}_{red}=\hbar\left(\begin{array}{ccc}
\omega_1 & \Delta_{12} & \Delta_{13}\\
\Delta_{12} & \omega_2  & 0\\
\Delta_{13} & 0 & \omega_3\\
\end{array}\right)\ .
\end{equation}
Still we will set $\hbar=1$. $\omega_1$, $\omega_2$ and $\omega_3$
are energy frequencies corresponding to the three states in
question. $\Delta_{12}$ and $\Delta_{13}$ are tunneling frequencies
between $|L0\rangle$ and the other two states respectively. They
mark the scales of the energy gaps at the two anticrossings.
$|R0\rangle$ and $|R1\rangle$ are not correlated.

Now we have a $3\times3$ density matrix
\begin{equation}
\hat\rho=\left(\begin{array}{ccc}
W_{11} & W_{12} & W_{13}\\
 W_{21} & W_{22}  & W_{23}\\
 W_{31} & W_{32} & W_{33}\\
\end{array}\right)\ .
\end{equation}
The calculation procedures will be the same as in Section \ref{3}.
We have changed the ratio $\Delta_{13}:\Delta_{12}$ and simulated
the interference in three cases, each of them having unique
characteristics. Now we explain the three cases in details.\\
\begin{figure}[hbpt]
\centering
\includegraphics[height=5cm]{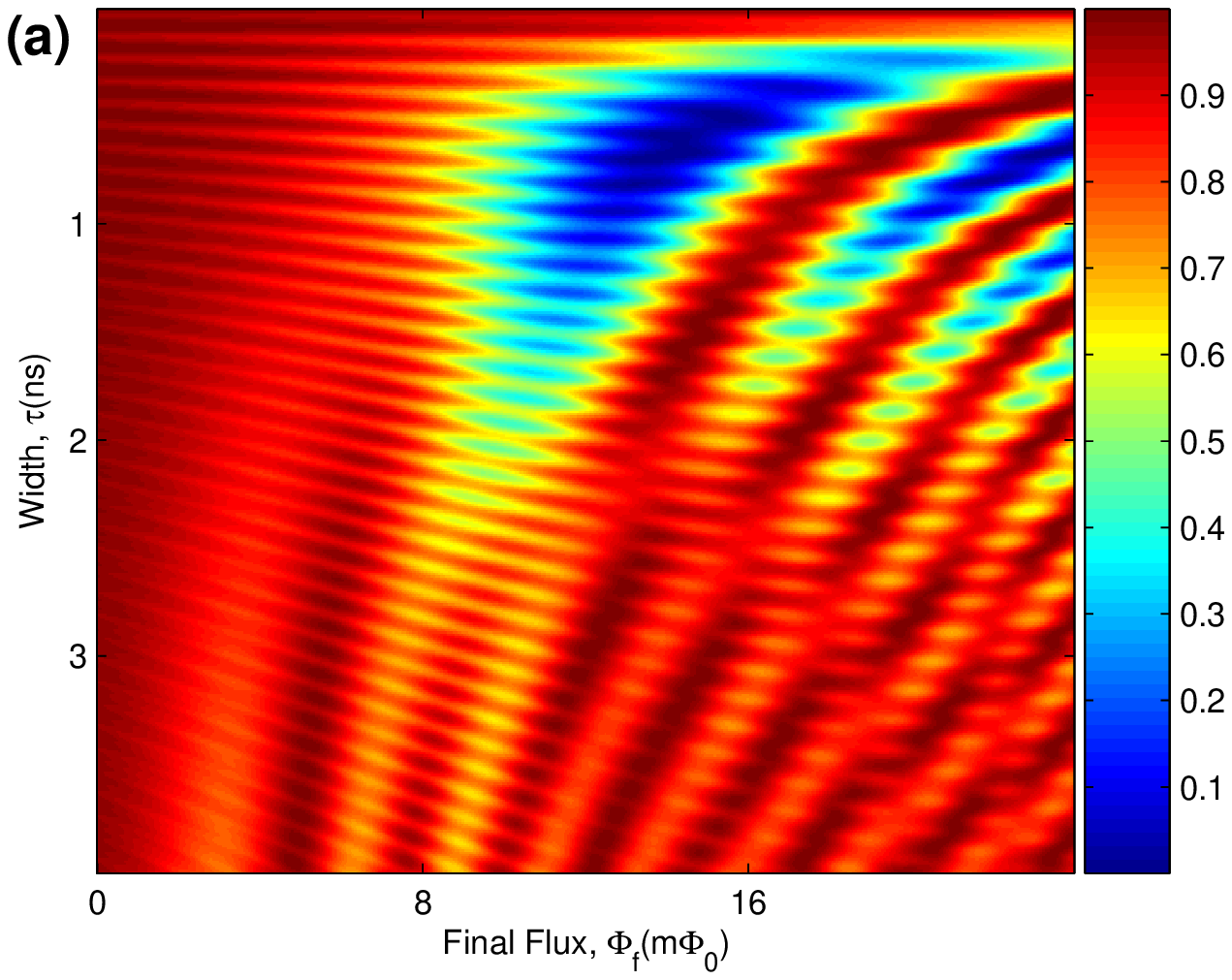}
\includegraphics[height=5cm]{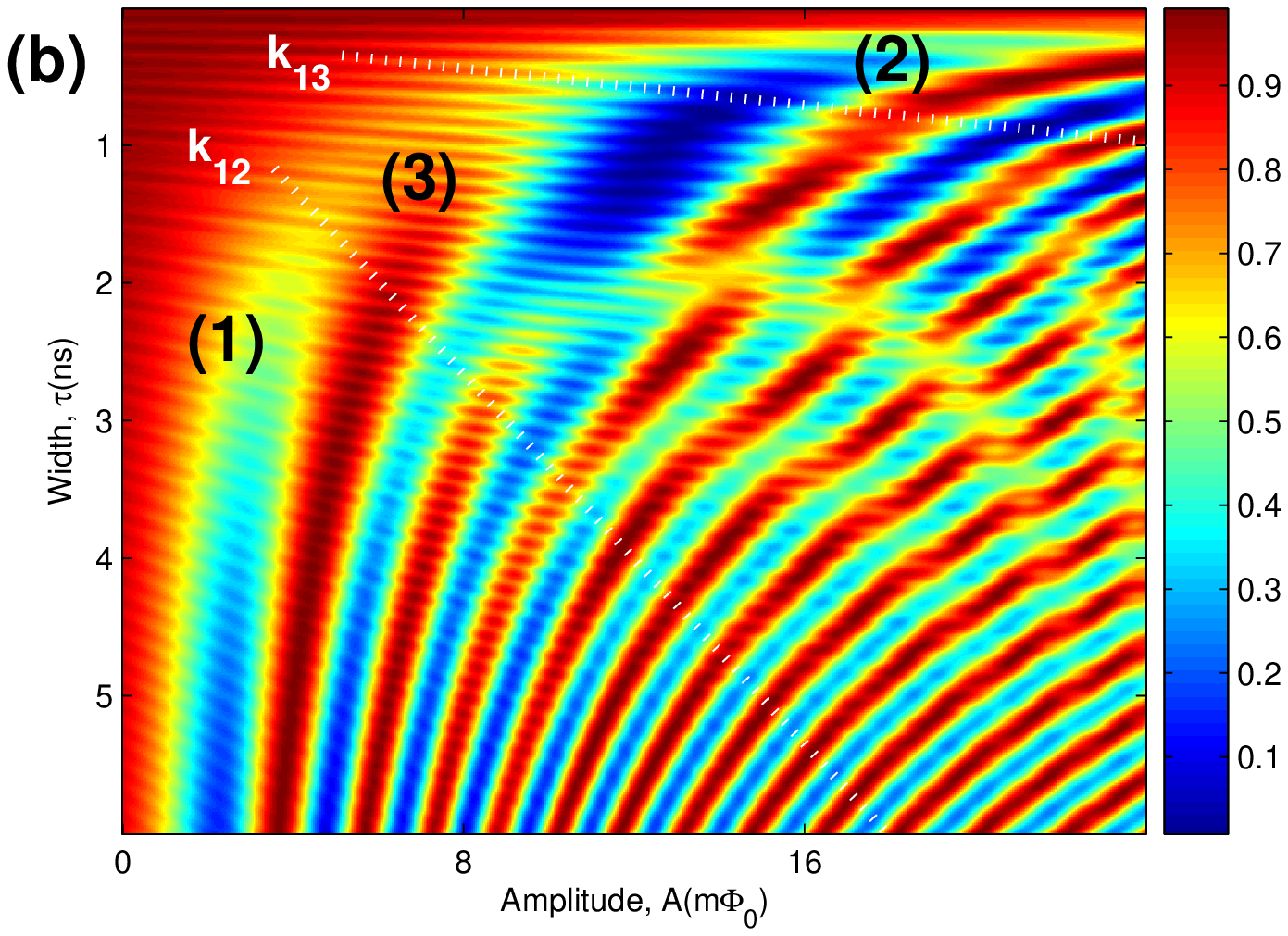}
\includegraphics[height=5cm]{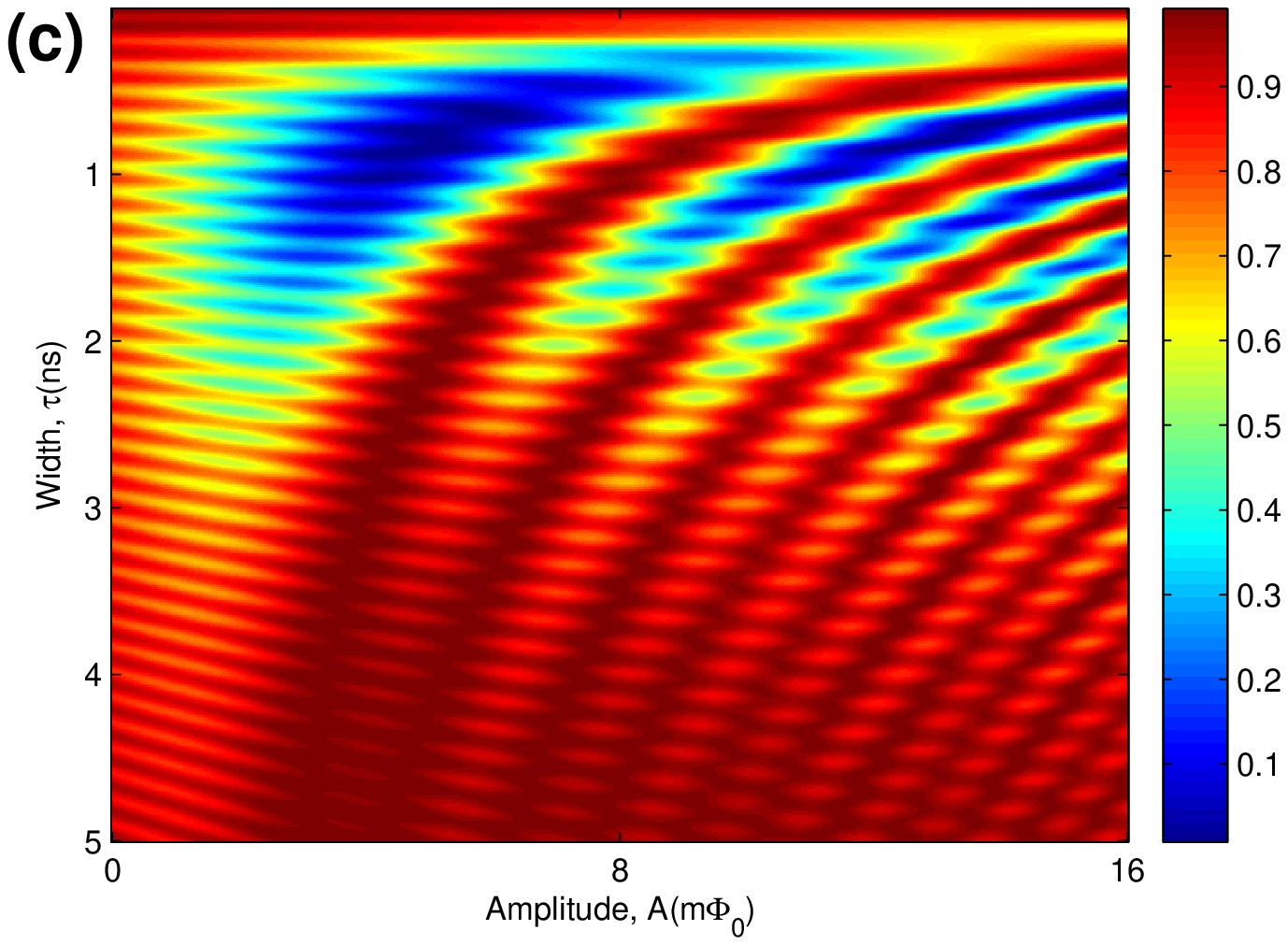}
\caption{\label{f2}(color online) Simulation results of LZS
interference with two anticrossings involved, marked by
$\Delta_{13}$ and $\Delta_{12}$ respectively. (a)In this case, we
have $\Delta_{13}:\Delta_{12}=10$. The interference fringes
primarily result from the bigger anticrossing, which can be located
by the left edge of the first fringe. Its parameters can be also
obtained by analyzing the pattern as a one-anticrossing case.
(b)Commensurate energy gaps are set. The ratio
$\Delta_{13}:\Delta_{12}$ is 4. The map can be divided into three
regions by two characteristic sweep rates $k_{12}$ and $k_{13}$,
shown with dot lines. In Region(3), distorted interference fringes
are observed, due to the cooperative effects of both anticrossings.
While in Regions(1) and (2), the pattern is similar to the
one-anticrossing case, since one of the anticrossings loses its
effect. Thus the information of the anticrossings can be inferred
from these two regions respectively. (c)Here
$\Delta_{13}:\Delta_{12}=1:4$, resulting in a pattern looking like
(a). The big anticrossing overwhelms the small one in most area. Its
information can be obtained as well.}
\end{figure}
\\
\textbf{Case 1} (Fig.(\ref{f2}a)): $\Delta_{13}=10$ GHz,
$\Delta_{12}=1$ GHz. We have the ratio $\Delta_{13}:\Delta_{12}$ as
large as 10 and obtain an interference pattern that looks similar to
the one-anticrossing case. The reason is that the first
anticrossing, marked by $\Delta_{12}$, is too small to affect the
results. There is no apparent oscillations when $0<\Phi_f<8$
m$\Phi_0$, since strong transitions keep occurring at the vicinities
as if there were no energy gaps. So, the interference fringes are
caused by the bigger anticrossing, whose location can be easily
inferred from the left edge of the first fringe. In this case we
find it at 8 m$\Phi_0$, in agreement with the spectrum in
Fig.(\ref{3spectrum}). Moreover, the parameters related to this
anticrossing can also be extracted, following the steps introduced
in Section \ref{3}. However, the information of the smaller
anticrossing cannot be obtained because
its effects are not revealed in this case. \\
\textbf{Case 2} (Fig.(\ref{f2}b)): $\Delta_{13}=8$ GHz,
$\Delta_{12}=2$ GHz. Then we lower the ratio
$\Delta_{13}:\Delta_{12}$ to 4 and then observe a significant
difference. The stripes are distorted and irregular when the second
anticrossing is reached. The distortion gets slighter as the signal
width increases and the pattern becomes close to that in the first
case. This can be anticipated because increasing width means slower
sweeping and thus gives rise to weaker transitions at the second
anticrossing. In the lower part of the map, the pattern resembles
that in the one-anticrossing situation, since all that contributes
is mainly the first one. Here we can obtain overall information of
the energy spectrum, because the two anticrossings are of
commensurate sizes that dual effects are observed. We will
particularly discuss this case later since it is of special interests.\\
\textbf{Case 3} (Fig.(\ref{f2}c)): $\Delta_{13}=2$ GHz,
$\Delta_{12}=8$ GHz. Now we set the ratio
$\Delta_{13}:\Delta_{12}=1/4$. We encounter a result similar to a
one-anticrossing case again. Like the first case, the reason is that
the contribution of the smaller anticrossing is rather trivial. It
is the first bigger anticrossing that primarily contributes in the
small-width region. In the figure we find clear interference fringes
from $\Phi_f=0$, where the first anticrossing is located. And the
details concerning this anticrossing can be obtained in the same way
as in Section \ref{3}.

In all the three figures there are very delicate ripples, which
result from the high precision of our calculation. While in
experiments, the ripples can hardly observed due to noises, decaying
or measurement limits.

Among the three cases we have discussed, the second is of special
interests and significance because it is the closest model to the
real situation. This is a typical multi-level system with two
anticrossings, on which our LZS spectroscopy method is applicable as
well. First of all, the locations of the anticrossings can be
inferred from the pattern. Similarly, the left edge of the first
fringe tells the location of the first anticrossing(at 0 m$\Phi_0$),
and the beginning of the distortion marks the second anticrossing
 (at 8 m$\Phi_0$). For further elaborations, we divide the map into
three regions with two dashed lines representing two characteristic
sweep rates $k_{1i}$, such that
$$\frac{2\pi\Delta_{1i}^2}{\hbar k_{1i}l}\simeq1 \ (\hbar=1,
i=2,3),$$ referring to Landau-Zener transition rate formula
\cite{Landau, Zener}. In Region(1) where $k\lesssim k_{12}$, the
pattern is mainly produced by the first anticrossing $\Delta_{12}$,
so we can calculate the energy slope and the value of $\Delta_{12}$.
Similarly, in Region(2) where $k\gtrsim k_{13}$, the information of
the second anticrossing $\Delta_{13}$ can be obtained. Region(3)
clearly demonstrates the cooperative effects of both anticrossings
and the transition from Region(1), dominated by $\Delta_{12}$, to
Region(2), dominated by $\Delta_{13}$. Along the direction of
increasing sweep rates, the distortion results in denser stripes. It
is because that the bigger anticrossing contributes more when
sweeping is faster, so the accumulated phase changes more rapidly,
according to Eq.(\ref{varphi}).

There have been some experimental results that qualitatively prove
our simulations. Recently, some researchers have reported that the
spectrum of an rf SQUID is modified by some spurious oscillators.
One of the significant works was carried out by Simmonds {\em et
al.} in 2004 \cite{tls}. The spurious oscillators, later interpreted
as two-level system (TLS), essentially and universally exist in
Josephson devices. They transform the energy structure of a phase
qubit and lead to multiple anticrossings by coupling with the qubit.
One of the latest works on such configuration is held by Sun {\em et
al.} \cite{sun}, who performed Landau-Zener interferometry in a
TLS-coupled phase qubit. The spectrum of the qubit shows two TLSs
coupled to the system, leading to two anticrossings on the left side
of $\Phi_{ext}=\Phi_0/2$. This two-anticrossing structure is similar
to the one we have discussed, although their origins are different.
Sun {\em et al.} used triangle pulses to drive the system, and
observed the interference fringes characterized by the distortion,
which has also been revealed in the second case of our simulations.
Notably, Shevchenko {\em et al.} also have made some deep
investigations into this issue  very recently \cite{Shevchenko}.

Additionally, we want to point out that the method of LZS
spectroscopy suggests a possibility of coherent manipulations on a
flux qubit. Since the whole process has a time scale of a very short
triangle pulse, it ensures that the operation is performed within
the decoherence time of the qubit. Moreover, with different sweep
rates, we can switch the two anticrossings on and off respectively.
The switching sweep rates are exactly the two we have used to
divide the map. We can interpret the three regions from an alternative view:\\
(1)$k\lesssim k_{12}$. Transitions occur mainly at the first
anticrossing but are ignorable at the second, as if the second anticrossing is switched off.\\
(2)$k\gtrsim k_{13}$. Now the first anticrossing is switched off
while the second is on, because transitions
occurring at the first anticrossing are very strong but moderate at the second.\\
(3)$k_{12}<k<k_{13}$. This situation corresponds to the region where
the two anticrossings are both switched on.

Therefore, the anticrossings act like tunable beam splitters with
transmission coefficients changing continuously from null to
unit\cite{sun}. By adjusting the sweeping rate carefully, we can
control the population in the excited states, supplying an
alternative method to manipulate the qubit.

\section {Conclusions \label{5}}
We propose a new method of measuring the energy spectrum of a
superconducting flux qubit. By sweeping the qubit through the energy
anticrossings with a linear triangle pulse, we obtain the LZS
interference patterns. Then by fitting the patterns with analytical
equations we can extract the information of the energy spectrum,
including the slope of the spectrum and the magnitudes of energy
gaps at anticrossings. We have demonstrated this method with
numerical simulations. It can be a more convenient and efficient
method with a precision over 90\%, especially in measuring a
spectrum with high energies, in which conventional frequency
spectroscopy has difficulties.

Another area where our method may be useful is in the adiabatic
quantum computation \cite{johansson, aqc}, in which microwaves are
totally unnecessary. If we use this linear spectroscopy to calibrate
the qubits, the measurement setup may be simplified and better
shielded from the extra noise, which is otherwise introduced from
microwave lines.

This method also can be applied to other quantum systems in which rf
field coupling may be too weak to generate detectable population
transitions, so we can try linear signals instead of rf fields to
perform spectroscopy and manipulations.

In addition, since the operation is done within the decoherence time
of a qubit, it can be utilized to realize coherent manipulations of
the qubit. We have also discussed the approach of controlling the
functions of the anticrossings. For our simulations qualitatively
agree with related experimental results, we think that our work
would stimulate more investigations in this field. Meanwhile, this
approach can be extended to multi-qubit systems, whose energy
spectra are of analogous structures. Therefore, the approach might
also shed light on large-scale controllable quantum computation in
future.

\begin{acknowledgements}Thanks to useful discussions with Xueda Wen. This work is partially supported by NSFC (10725415) and the State
Key Program for Basic Research of China (2006CB921801).
\end{acknowledgements}

\bibliography{LZSI}

\begin{thebibliography}{37}
\expandafter\ifx\csname natexlab\endcsname\relax\def\natexlab#1{#1}\fi
\expandafter\ifx\csname bibnamefont\endcsname\relax
  \def\bibnamefont#1{#1}\fi
\expandafter\ifx\csname bibfnamefont\endcsname\relax
  \def\bibfnamefont#1{#1}\fi
\expandafter\ifx\csname citenamefont\endcsname\relax
  \def\citenamefont#1{#1}\fi
\expandafter\ifx\csname url\endcsname\relax
  \def\url#1{\texttt{#1}}\fi
\expandafter\ifx\csname urlprefix\endcsname\relax\def\urlprefix{URL }\fi
\providecommand{\bibinfo}[2]{#2}
\providecommand{\eprint}[2][]{\url{#2}}

\bibitem[{\citenamefont{Makhlin et~al.}(2001)\citenamefont{Makhlin, {G.
  Sch\"{o}n}, and Shnirman}}]{qrev}
\bibinfo{author}{\bibfnamefont{Y.}~\bibnamefont{Makhlin}},
  \bibinfo{author}{\bibnamefont{{G. Sch\"{o}n}}}, \bibnamefont{and}
  \bibinfo{author}{\bibfnamefont{A.}~\bibnamefont{Shnirman}},
  \bibinfo{journal}{\rmp} \textbf{\bibinfo{volume}{73}}, \bibinfo{pages}{357}
  (\bibinfo{year}{2001}).

\bibitem[{\citenamefont{Caldeira and Leggett}(1981)}]{cal_leg}
\bibinfo{author}{\bibfnamefont{A.~O.} \bibnamefont{Caldeira}} \bibnamefont{and}
  \bibinfo{author}{\bibfnamefont{A.~J.} \bibnamefont{Leggett}},
  \bibinfo{journal}{\rmp} \textbf{\bibinfo{volume}{46}}, \bibinfo{pages}{4}
  (\bibinfo{year}{1981}).

\bibitem[{\citenamefont{Devoret et~al.}(1985)\citenamefont{Devoret, Martinis,
  and Clarke}}]{tunneling}
\bibinfo{author}{\bibfnamefont{M.~H.} \bibnamefont{Devoret}},
  \bibinfo{author}{\bibfnamefont{J.~M.} \bibnamefont{Martinis}},
  \bibnamefont{and} \bibinfo{author}{\bibfnamefont{J.}~\bibnamefont{Clarke}},
  \bibinfo{journal}{\prl} \textbf{\bibinfo{volume}{55}}, \bibinfo{pages}{18}
  (\bibinfo{year}{1985}).

\bibitem[{\citenamefont{Han et~al.}(2001)\citenamefont{Han, Yu, Chu, Chu, and
  Wang}}]{han2001}
\bibinfo{author}{\bibfnamefont{S.}~\bibnamefont{Han}},
  \bibinfo{author}{\bibfnamefont{Y.}~\bibnamefont{Yu}},
  \bibinfo{author}{\bibfnamefont{X.}~\bibnamefont{Chu}},
  \bibinfo{author}{\bibfnamefont{S.-I.} \bibnamefont{Chu}}, \bibnamefont{and}
  \bibinfo{author}{\bibfnamefont{Z.}~\bibnamefont{Wang}},
  \bibinfo{journal}{Science} \textbf{\bibinfo{volume}{293}},
  \bibinfo{pages}{1457} (\bibinfo{year}{2001}).

\bibitem[{\citenamefont{Martinis et~al.}(1985)\citenamefont{Martinis, Devoret,
  and Clarke}}]{martinis1985}
\bibinfo{author}{\bibfnamefont{J.~M.} \bibnamefont{Martinis}},
  \bibinfo{author}{\bibfnamefont{M.~H.} \bibnamefont{Devoret}},
  \bibnamefont{and} \bibinfo{author}{\bibfnamefont{J.}~\bibnamefont{Clarke}},
  \bibinfo{journal}{\prl} \textbf{\bibinfo{volume}{55}}, \bibinfo{pages}{15}
  (\bibinfo{year}{1985}).

\bibitem[{\citenamefont{Nakamura et~al.}(1999)\citenamefont{Nakamura, Pashkin,
  and Tsai}}]{nakamura1999}
\bibinfo{author}{\bibfnamefont{Y.}~\bibnamefont{Nakamura}},
  \bibinfo{author}{\bibfnamefont{Y.~A.} \bibnamefont{Pashkin}},
  \bibnamefont{and} \bibinfo{author}{\bibfnamefont{J.~S.} \bibnamefont{Tsai}},
  \bibinfo{journal}{Nature} \textbf{\bibinfo{volume}{398}},
  \bibinfo{pages}{786} (\bibinfo{year}{1999}).

\bibitem[{\citenamefont{Friedman et~al.}(2000)\citenamefont{Friedman, Patel,
  Chen, Tolpygo, and Lukens}}]{sup}
\bibinfo{author}{\bibfnamefont{J.~R.} \bibnamefont{Friedman}},
  \bibinfo{author}{\bibfnamefont{V.}~\bibnamefont{Patel}},
  \bibinfo{author}{\bibfnamefont{W.}~\bibnamefont{Chen}},
  \bibinfo{author}{\bibfnamefont{S.~K.} \bibnamefont{Tolpygo}},
  \bibnamefont{and} \bibinfo{author}{\bibfnamefont{J.~E.}
  \bibnamefont{Lukens}}, \bibinfo{journal}{Nature}
  \textbf{\bibinfo{volume}{406}}, \bibinfo{pages}{43} (\bibinfo{year}{2000}).

\bibitem[{\citenamefont{van~der Wal et~al.}(2000)\citenamefont{van~der Wal, ter
  Haar, Wilhelm, Schouten, Harmans, Orlando, Lloyd, and Mooij}}]{suppcq}
\bibinfo{author}{\bibfnamefont{C.~H.} \bibnamefont{van~der Wal}},
  \bibinfo{author}{\bibfnamefont{A.~C.~J.} \bibnamefont{ter Haar}},
  \bibinfo{author}{\bibfnamefont{F.~K.} \bibnamefont{Wilhelm}},
  \bibinfo{author}{\bibfnamefont{R.~N.} \bibnamefont{Schouten}},
  \bibinfo{author}{\bibfnamefont{C.~J. P.~M.} \bibnamefont{Harmans}},
  \bibinfo{author}{\bibfnamefont{T.~P.} \bibnamefont{Orlando}},
  \bibinfo{author}{\bibfnamefont{S.}~\bibnamefont{Lloyd}}, \bibnamefont{and}
  \bibinfo{author}{\bibfnamefont{J.~E.} \bibnamefont{Mooij}},
  \bibinfo{journal}{Science} \textbf{\bibinfo{volume}{290}},
  \bibinfo{pages}{773} (\bibinfo{year}{2000}).

\bibitem[{\citenamefont{Chiorescu et~al.}(2003)\citenamefont{Chiorescu,
  Nakamura, Harmans, and Mooij}}]{cohdyn}
\bibinfo{author}{\bibfnamefont{I.}~\bibnamefont{Chiorescu}},
  \bibinfo{author}{\bibfnamefont{Y.}~\bibnamefont{Nakamura}},
  \bibinfo{author}{\bibfnamefont{C.~J. P.~M.} \bibnamefont{Harmans}},
  \bibnamefont{and} \bibinfo{author}{\bibfnamefont{J.~E.} \bibnamefont{Mooij}},
  \bibinfo{journal}{Science} \textbf{\bibinfo{volume}{299}},
  \bibinfo{pages}{1869} (\bibinfo{year}{2003}).

\bibitem[{\citenamefont{Martinis et~al.}(2003)\citenamefont{Martinis, Nam,
  Aumentado, and Urbina}}]{rabi}
\bibinfo{author}{\bibfnamefont{J.~M.} \bibnamefont{Martinis}},
  \bibinfo{author}{\bibfnamefont{S.}~\bibnamefont{Nam}},
  \bibinfo{author}{\bibfnamefont{J.}~\bibnamefont{Aumentado}},
  \bibnamefont{and} \bibinfo{author}{\bibfnamefont{C.}~\bibnamefont{Urbina}},
  \bibinfo{journal}{\prl} \textbf{\bibinfo{volume}{89}}, \bibinfo{pages}{11}
  (\bibinfo{year}{2003}).

\bibitem[{\citenamefont{Yu et~al.}(2002)\citenamefont{Yu, Han, Chu, Chu, and
  Wang}}]{cohosc}
\bibinfo{author}{\bibfnamefont{Y.}~\bibnamefont{Yu}},
  \bibinfo{author}{\bibfnamefont{S.}~\bibnamefont{Han}},
  \bibinfo{author}{\bibfnamefont{X.}~\bibnamefont{Chu}},
  \bibinfo{author}{\bibfnamefont{S.-I.} \bibnamefont{Chu}}, \bibnamefont{and}
  \bibinfo{author}{\bibfnamefont{Z.}~\bibnamefont{Wang}},
  \bibinfo{journal}{Science} \textbf{\bibinfo{volume}{296}},
  \bibinfo{pages}{889} (\bibinfo{year}{2002}).

\bibitem[{\citenamefont{Nielsen and Chuang}(2000)}]{qcqi}
\bibinfo{author}{\bibfnamefont{M.~A.} \bibnamefont{Nielsen}} \bibnamefont{and}
  \bibinfo{author}{\bibfnamefont{I.~L.} \bibnamefont{Chuang}},
  \emph{\bibinfo{title}{Quantum Computation and Quantum Information}}
  (\bibinfo{publisher}{Cambridge University Press}, \bibinfo{year}{2000}).

\bibitem[{\citenamefont{Mooij et~al.}(1999)\citenamefont{Mooij, Orlando,
  Levitov, Tian, van~der Wal, and Lloyd}}]{pcqubit}
\bibinfo{author}{\bibfnamefont{J.~E.} \bibnamefont{Mooij}},
  \bibinfo{author}{\bibfnamefont{T.~P.} \bibnamefont{Orlando}},
  \bibinfo{author}{\bibfnamefont{L.}~\bibnamefont{Levitov}},
  \bibinfo{author}{\bibfnamefont{L.}~\bibnamefont{Tian}},
  \bibinfo{author}{\bibfnamefont{C.~H.} \bibnamefont{van~der Wal}},
  \bibnamefont{and} \bibinfo{author}{\bibfnamefont{S.}~\bibnamefont{Lloyd}},
  \bibinfo{journal}{Science} \textbf{\bibinfo{volume}{285}},
  \bibinfo{pages}{1036} (\bibinfo{year}{1999}).

\bibitem[{\citenamefont{You and Nori}(2005)}]{you}
\bibinfo{author}{\bibfnamefont{J.}~\bibnamefont{You}} \bibnamefont{and}
  \bibinfo{author}{\bibfnamefont{F.}~\bibnamefont{Nori}},
  \bibinfo{journal}{Phys. Today} \textbf{\bibinfo{volume}{58}},
  \bibinfo{pages}{42} (\bibinfo{year}{2005}).

\bibitem[{\citenamefont{Pashkin et~al.}(2003)\citenamefont{Pashkin, Yamamoto,
  Astafiev, Nakamura, Averin, and Tsai}}]{pashkin}
\bibinfo{author}{\bibfnamefont{Y.~A.} \bibnamefont{Pashkin}},
  \bibinfo{author}{\bibfnamefont{T.}~\bibnamefont{Yamamoto}},
  \bibinfo{author}{\bibfnamefont{O.}~\bibnamefont{Astafiev}},
  \bibinfo{author}{\bibfnamefont{Y.}~\bibnamefont{Nakamura}},
  \bibinfo{author}{\bibfnamefont{D.~V.} \bibnamefont{Averin}},
  \bibnamefont{and} \bibinfo{author}{\bibfnamefont{J.~S.} \bibnamefont{Tsai}},
  \bibinfo{journal}{Nature} \textbf{\bibinfo{volume}{421}},
  \bibinfo{pages}{823} (\bibinfo{year}{2003}).

\bibitem[{\citenamefont{Berkley et~al.}(2003)\citenamefont{Berkley, Xu, Ramos,
  Gubrud, Strauch, Johnson, Anderson, Dragt, Lobb, and
  Wellstood}}]{entangled2qubits}
\bibinfo{author}{\bibfnamefont{A.~J.} \bibnamefont{Berkley}},
  \bibinfo{author}{\bibfnamefont{H.}~\bibnamefont{Xu}},
  \bibinfo{author}{\bibfnamefont{R.~C.} \bibnamefont{Ramos}},
  \bibinfo{author}{\bibfnamefont{M.~A.} \bibnamefont{Gubrud}},
  \bibinfo{author}{\bibfnamefont{F.~W.} \bibnamefont{Strauch}},
  \bibinfo{author}{\bibfnamefont{P.~R.} \bibnamefont{Johnson}},
  \bibinfo{author}{\bibfnamefont{J.~R.} \bibnamefont{Anderson}},
  \bibinfo{author}{\bibfnamefont{A.~J.} \bibnamefont{Dragt}},
  \bibinfo{author}{\bibfnamefont{C.~J.} \bibnamefont{Lobb}}, \bibnamefont{and}
  \bibinfo{author}{\bibfnamefont{F.~C.} \bibnamefont{Wellstood}},
  \bibinfo{journal}{Science} \textbf{\bibinfo{volume}{300}},
  \bibinfo{pages}{1548} (\bibinfo{year}{2003}).

\bibitem[{\citenamefont{Hime et~al.}(2006)\citenamefont{Hime, Reichardt,
  Plourde, Robertson, Wu, Ustinov, and Clarke}}]{controlcoup}
\bibinfo{author}{\bibfnamefont{T.}~\bibnamefont{Hime}},
  \bibinfo{author}{\bibfnamefont{P.~A.} \bibnamefont{Reichardt}},
  \bibinfo{author}{\bibfnamefont{B.~L.~T.} \bibnamefont{Plourde}},
  \bibinfo{author}{\bibfnamefont{T.~L.} \bibnamefont{Robertson}},
  \bibinfo{author}{\bibfnamefont{C.-E.} \bibnamefont{Wu}},
  \bibinfo{author}{\bibfnamefont{A.~V.} \bibnamefont{Ustinov}},
  \bibnamefont{and} \bibinfo{author}{\bibfnamefont{J.}~\bibnamefont{Clarke}},
  \bibinfo{journal}{Science} \textbf{\bibinfo{volume}{314}},
  \bibinfo{pages}{1427} (\bibinfo{year}{2006}).

\bibitem[{\citenamefont{DiCarlo et~al.}(2009)\citenamefont{DiCarlo, Chow,
  Gambetta, Bishop, Johnson, Schuster, Majer, Blais, Frunzio, Girvin
  et~al.}}]{twoqubit}
\bibinfo{author}{\bibfnamefont{L.}~\bibnamefont{DiCarlo}},
  \bibinfo{author}{\bibfnamefont{J.~M.} \bibnamefont{Chow}},
  \bibinfo{author}{\bibfnamefont{J.~M.} \bibnamefont{Gambetta}},
  \bibinfo{author}{\bibfnamefont{L.~S.} \bibnamefont{Bishop}},
  \bibinfo{author}{\bibfnamefont{B.~R.} \bibnamefont{Johnson}},
  \bibinfo{author}{\bibfnamefont{D.~I.} \bibnamefont{Schuster}},
  \bibinfo{author}{\bibfnamefont{J.}~\bibnamefont{Majer}},
  \bibinfo{author}{\bibfnamefont{A.}~\bibnamefont{Blais}},
  \bibinfo{author}{\bibfnamefont{L.}~\bibnamefont{Frunzio}},
  \bibinfo{author}{\bibfnamefont{S.~M.} \bibnamefont{Girvin}},
  \bibnamefont{et~al.}, \bibinfo{journal}{Nature}
  \textbf{\bibinfo{volume}{460}}, \bibinfo{pages}{240} (\bibinfo{year}{2009}).

\bibitem[{\citenamefont{Yang et~al.}(2010)\citenamefont{Yang, Liu, , and
  Nori}}]{multiqubit}
\bibinfo{author}{\bibfnamefont{C.-P.} \bibnamefont{Yang}},
  \bibinfo{author}{\bibfnamefont{Y.-X.} \bibnamefont{Liu}}, , \bibnamefont{and}
  \bibinfo{author}{\bibfnamefont{F.}~\bibnamefont{Nori}},
  \bibinfo{journal}{\pra} \textbf{\bibinfo{volume}{81}},
  \bibinfo{pages}{062323} (\bibinfo{year}{2010}).

\bibitem[{\citenamefont{Schawlow}(1982)}]{a1}
\bibinfo{author}{\bibfnamefont{A.~L.} \bibnamefont{Schawlow}},
  \bibinfo{journal}{\rmp} \textbf{\bibinfo{volume}{54}}, \bibinfo{pages}{697}
  (\bibinfo{year}{1982}).

\bibitem[{\citenamefont{Thompson}(1985)}]{a2}
\bibinfo{author}{\bibfnamefont{R.~C.} \bibnamefont{Thompson}},
  \bibinfo{journal}{Rep. Prog. Phys.} \textbf{\bibinfo{volume}{48}},
  \bibinfo{pages}{531} (\bibinfo{year}{1985}).

\bibitem[{\citenamefont{Collin}(2001)}]{ieee}
\bibinfo{author}{\bibfnamefont{R.~E.} \bibnamefont{Collin}},
  \emph{\bibinfo{title}{Foundations for Microwave Engineering}}
  (\bibinfo{publisher}{Wiley-IEEE}, \bibinfo{year}{2001}).

\bibitem[{\citenamefont{Clarke et~al.}(1988)\citenamefont{Clarke, Cleland,
  Devoret, Esteve, and Martinis}}]{photonat}
\bibinfo{author}{\bibfnamefont{J.}~\bibnamefont{Clarke}},
  \bibinfo{author}{\bibfnamefont{A.~N.} \bibnamefont{Cleland}},
  \bibinfo{author}{\bibfnamefont{M.~H.} \bibnamefont{Devoret}},
  \bibinfo{author}{\bibfnamefont{D.}~\bibnamefont{Esteve}}, \bibnamefont{and}
  \bibinfo{author}{\bibfnamefont{J.~H.} \bibnamefont{Martinis}},
  \bibinfo{journal}{Science} \textbf{\bibinfo{volume}{239}},
  \bibinfo{pages}{992} (\bibinfo{year}{1988}).

\bibitem[{\citenamefont{Berns et~al.}(2008)\citenamefont{Berns, Rudner,
  Valenzuela, Berggren, Oliver, Levitov, and Orlando}}]{as}
\bibinfo{author}{\bibfnamefont{D.~M.} \bibnamefont{Berns}},
  \bibinfo{author}{\bibfnamefont{M.~S.} \bibnamefont{Rudner}},
  \bibinfo{author}{\bibfnamefont{S.~O.} \bibnamefont{Valenzuela}},
  \bibinfo{author}{\bibfnamefont{K.~K.} \bibnamefont{Berggren}},
  \bibinfo{author}{\bibfnamefont{W.~D.} \bibnamefont{Oliver}},
  \bibinfo{author}{\bibfnamefont{L.~S.} \bibnamefont{Levitov}},
  \bibnamefont{and} \bibinfo{author}{\bibfnamefont{T.~P.}
  \bibnamefont{Orlando}}, \bibinfo{journal}{Nature}
  \textbf{\bibinfo{volume}{455}}, \bibinfo{pages}{51} (\bibinfo{year}{2008}).

\bibitem[{\citenamefont{Bylander et~al.}(2009)\citenamefont{Bylander, Rudner,
  Shytov, Valenzuela, Berns, Berggren, Levitov, and Oliver}}]{oliver2009}
\bibinfo{author}{\bibfnamefont{J.}~\bibnamefont{Bylander}},
  \bibinfo{author}{\bibfnamefont{M.~S.} \bibnamefont{Rudner}},
  \bibinfo{author}{\bibfnamefont{A.~V.} \bibnamefont{Shytov}},
  \bibinfo{author}{\bibfnamefont{S.~O.} \bibnamefont{Valenzuela}},
  \bibinfo{author}{\bibfnamefont{D.~M.} \bibnamefont{Berns}},
  \bibinfo{author}{\bibfnamefont{K.~K.} \bibnamefont{Berggren}},
  \bibinfo{author}{\bibfnamefont{L.~S.} \bibnamefont{Levitov}},
  \bibnamefont{and} \bibinfo{author}{\bibfnamefont{W.~D.}
  \bibnamefont{Oliver}}, \bibinfo{journal}{\prb} \textbf{\bibinfo{volume}{80}},
  \bibinfo{pages}{220506} (\bibinfo{year}{2009}).

\bibitem[{\citenamefont{Shytov et~al.}(2003)\citenamefont{Shytov, Ivanov, and
  Feigel¡¯man}}]{LZ4q}
\bibinfo{author}{\bibfnamefont{A.~V.} \bibnamefont{Shytov}},
  \bibinfo{author}{\bibfnamefont{D.~A.} \bibnamefont{Ivanov}},
  \bibnamefont{and} \bibinfo{author}{\bibfnamefont{M.~V.}
  \bibnamefont{Feigel¡¯man}}, \bibinfo{journal}{Eur. Phys. J. B}
  \textbf{\bibinfo{volume}{36}}, \bibinfo{pages}{263} (\bibinfo{year}{2003}).

\bibitem[{\citenamefont{Ashhab et~al.}(2007)\citenamefont{Ashhab, Johansson,
  Zagoskin, and Nori}}]{la4tls}
\bibinfo{author}{\bibfnamefont{S.}~\bibnamefont{Ashhab}},
  \bibinfo{author}{\bibfnamefont{J.~R.} \bibnamefont{Johansson}},
  \bibinfo{author}{\bibfnamefont{A.~M.} \bibnamefont{Zagoskin}},
  \bibnamefont{and} \bibinfo{author}{\bibfnamefont{F.}~\bibnamefont{Nori}},
  \bibinfo{journal}{\pra} \textbf{\bibinfo{volume}{75}},
  \bibinfo{pages}{063414} (\bibinfo{year}{2007}).

\bibitem[{\citenamefont{Wen and Yu}(2009)}]{multilevels}
\bibinfo{author}{\bibfnamefont{X.}~\bibnamefont{Wen}} \bibnamefont{and}
  \bibinfo{author}{\bibfnamefont{Y.}~\bibnamefont{Yu}}, \bibinfo{journal}{\prb}
  \textbf{\bibinfo{volume}{79}}, \bibinfo{pages}{094529}
  (\bibinfo{year}{2009}).

\bibitem[{\citenamefont{Oliver et~al.}(2005)\citenamefont{Oliver, Yu, Lee,
  Berggren, Levitov, and Orlando}}]{MZI}
\bibinfo{author}{\bibfnamefont{W.~D.} \bibnamefont{Oliver}},
  \bibinfo{author}{\bibfnamefont{Y.}~\bibnamefont{Yu}},
  \bibinfo{author}{\bibfnamefont{J.~C.} \bibnamefont{Lee}},
  \bibinfo{author}{\bibfnamefont{K.~K.} \bibnamefont{Berggren}},
  \bibinfo{author}{\bibfnamefont{L.~S.} \bibnamefont{Levitov}},
  \bibnamefont{and} \bibinfo{author}{\bibfnamefont{T.~P.}
  \bibnamefont{Orlando}}, \bibinfo{journal}{Science}
  \textbf{\bibinfo{volume}{310}}, \bibinfo{pages}{1653} (\bibinfo{year}{2005}).

\bibitem[{\citenamefont{Johansson et~al.}(2009)\citenamefont{Johansson, Amin,
  Berkley, Bunyk, Choi, Harris, Johnson, Lanting, Lloyd, and Rose}}]{johansson}
\bibinfo{author}{\bibfnamefont{J.}~\bibnamefont{Johansson}},
  \bibinfo{author}{\bibfnamefont{M.}~\bibnamefont{Amin}},
  \bibinfo{author}{\bibfnamefont{A.}~\bibnamefont{Berkley}},
  \bibinfo{author}{\bibfnamefont{P.}~\bibnamefont{Bunyk}},
  \bibinfo{author}{\bibfnamefont{V.}~\bibnamefont{Choi}},
  \bibinfo{author}{\bibfnamefont{R.}~\bibnamefont{Harris}},
  \bibinfo{author}{\bibfnamefont{M.}~\bibnamefont{Johnson}},
  \bibinfo{author}{\bibfnamefont{T.}~\bibnamefont{Lanting}},
  \bibinfo{author}{\bibfnamefont{S.}~\bibnamefont{Lloyd}}, \bibnamefont{and}
  \bibinfo{author}{\bibfnamefont{G.}~\bibnamefont{Rose}},
  \bibinfo{journal}{\prb} \textbf{\bibinfo{volume}{80}},
  \bibinfo{pages}{012507} (\bibinfo{year}{2009}).

\bibitem[{\citenamefont{Zener}(1932)}]{Zener}
\bibinfo{author}{\bibfnamefont{C.}~\bibnamefont{Zener}},
  \bibinfo{journal}{Proc. R. Soc. London, Ser. A}
  \textbf{\bibinfo{volume}{137}}, \bibinfo{pages}{696} (\bibinfo{year}{1932}).

\bibitem[{\citenamefont{Xu}(2010)}]{Xu}
\bibinfo{author}{\bibfnamefont{S.}~\bibnamefont{Xu}}, \bibinfo{journal}{B.S.
  Thesis {``Landau-Zener-St\"{u}ckelberg Interference in Josephson Junction
  Devices"}}  (\bibinfo{year}{2010}).

\bibitem[{\citenamefont{Landau}(1932)}]{Landau}
\bibinfo{author}{\bibfnamefont{L.~D.} \bibnamefont{Landau}},
  \bibinfo{journal}{Phys. Z. Sowjetunion} \textbf{\bibinfo{volume}{1}},
  \bibinfo{pages}{89} (\bibinfo{year}{1932}).

\bibitem[{\citenamefont{Simmonds et~al.}(2004)\citenamefont{Simmonds, Lang,
  Hite, Nam, Pappas, and Martinis}}]{tls}
\bibinfo{author}{\bibfnamefont{R.~W.} \bibnamefont{Simmonds}},
  \bibinfo{author}{\bibfnamefont{K.~M.} \bibnamefont{Lang}},
  \bibinfo{author}{\bibfnamefont{D.~A.} \bibnamefont{Hite}},
  \bibinfo{author}{\bibfnamefont{S.}~\bibnamefont{Nam}},
  \bibinfo{author}{\bibfnamefont{D.~P.} \bibnamefont{Pappas}},
  \bibnamefont{and} \bibinfo{author}{\bibfnamefont{J.~M.}
  \bibnamefont{Martinis}}, \bibinfo{journal}{\prl}
  \textbf{\bibinfo{volume}{93}}, \bibinfo{pages}{7} (\bibinfo{year}{2004}).

\bibitem[{\citenamefont{Sun et~al.}(2010)\citenamefont{Sun, Wen, Mao, Chen, Yu,
  Wu, and Han}}]{sun}
\bibinfo{author}{\bibfnamefont{G.}~\bibnamefont{Sun}},
  \bibinfo{author}{\bibfnamefont{X.}~\bibnamefont{Wen}},
  \bibinfo{author}{\bibfnamefont{B.}~\bibnamefont{Mao}},
  \bibinfo{author}{\bibfnamefont{J.}~\bibnamefont{Chen}},
  \bibinfo{author}{\bibfnamefont{Y.}~\bibnamefont{Yu}},
  \bibinfo{author}{\bibfnamefont{P.}~\bibnamefont{Wu}}, \bibnamefont{and}
  \bibinfo{author}{\bibfnamefont{S.}~\bibnamefont{Han}},
  \bibinfo{journal}{arXiv:1004.4657v2 [cond-mat.supr-con]}
  (\bibinfo{year}{2010}).

\bibitem[{\citenamefont{Shevchenko et~al.}(2010)\citenamefont{Shevchenko,
  Ashhabb, and Nori}}]{Shevchenko}
\bibinfo{author}{\bibfnamefont{S.}~\bibnamefont{Shevchenko}},
  \bibinfo{author}{\bibfnamefont{S.}~\bibnamefont{Ashhabb}}, \bibnamefont{and}
  \bibinfo{author}{\bibfnamefont{F.}~\bibnamefont{Nori}},
  \bibinfo{journal}{Phys. Rep.} \textbf{\bibinfo{volume}{482}},
  \bibinfo{pages}{1} (\bibinfo{year}{2010}).

\bibitem[{\citenamefont{Farhi et~al.}(2001)\citenamefont{Farhi, Goldstone,
  Gutmann, Lapan, Lundgren, and Preda}}]{aqc}
\bibinfo{author}{\bibfnamefont{E.}~\bibnamefont{Farhi}},
  \bibinfo{author}{\bibfnamefont{J.}~\bibnamefont{Goldstone}},
  \bibinfo{author}{\bibfnamefont{S.}~\bibnamefont{Gutmann}},
  \bibinfo{author}{\bibfnamefont{J.}~\bibnamefont{Lapan}},
  \bibinfo{author}{\bibfnamefont{A.}~\bibnamefont{Lundgren}}, \bibnamefont{and}
  \bibinfo{author}{\bibfnamefont{D.}~\bibnamefont{Preda}},
  \bibinfo{journal}{Science} \textbf{\bibinfo{volume}{292}},
  \bibinfo{pages}{472} (\bibinfo{year}{2001}).

\end{thebibliography}

\end{document}